\documentstyle[12pt]{article}
\begin{document}
\thispagestyle{empty}
\begin{center}
\LARGE \tt \bf{Teleparallel Textures}
\end{center}
\vspace{2.5cm}
\begin{center}{\large L.C. Garcia de Andrade\footnote{Departamento de F\'{\i}sica Teorica - UERJ.
Rua S\~{a}o Fco. Xavier 524, Rio de Janeiro, RJ
Maracan\~{a}, CEP:20550-003 , Brasil.
{e-mail: garcia@dft.if.uerj.br}}}
\end{center}
\vspace{2.0cm}
\begin{abstract}
An example of a teleparallel texture is given by an appropriate choice of torsion components in the tetrad frame.In the light cone limit the metric is not globally Euclidean and the spherical angles depend on torsion similarly to what happens in cosmic string space-times.In this limit torsion produces a force which decays with $r^{-3}$. 
\end{abstract}
\vspace{2.0cm}
\newpage
Recently a renewed interest in teleparallel theories of gravity \cite{1} has emerged with the semi-teleparallel theory of gravity introduced by C.Koehler \cite{2}.In the past a controversy on the non-predictable behaviour of torsion for certain Lagrangians in the metric-teleparallel theories of gravity has been raised by Kopczynski \cite{3} and apparently solved by Nester \cite{4}.More recently Cartan torsion has been successfully applied to topological defects \cite{5,6,7}.
From the well-known topological defects \cite{8} (monopoles,cosmic  strings,global textures and domain walls) certainly along with cosmic strings,textures are the most intesting one for the cosmic structure for the formation of galaxies.In this Letter we present an example of a teleparallel texture as an exact solution for the telaparallel theory of gravity where to simplify matters we make a suitable choice of torsion components as in the tetrad frame of differential forms.The physical and global properties of the metric are also discussed.To find our solution we use as an ansatz a metric discovered by Notzold \cite{8} which describes a Riemannian texture.
This metric is given by
\begin{equation}
ds^{2}=dt^{2}-dr^{2}-(1+{\omega})r^{2}d{\Omega}^{2}
\label{1}
\end{equation}
where $d{\Omega}^{2}=d{\theta}^{2}+sin^{2}{\theta}d{\phi}^{2}$ and ${\omega}={\omega}(t,r)$.In the case of Notzold solution ${\omega}=-2{\epsilon}[1+{\frac{t}{r}}cot^{-1}(\frac{-t}{r})]$,where ${\epsilon}=8{\pi}G{\eta}^{2}$ but of course here the solution would be of different nature.In orthonormal tetrads ${a=0,1,2,3}$ are the tetradic indices where the orthonormal metric is given by ${\eta}_{ab}=diag(1,-1,-1,-1)$, the above metric can be expressed as 
\begin{equation}
ds^{2}={\eta}_{ab}dx^{a}dx^{b} 
\label{2}
\end{equation}
with the 1-form basis ${e^{a}}$ given by $e^{0}=dt$,$e^{1}=dr$,$e^{2}=(1+{\omega})^{\frac{1}{2}}rd{\theta}$ and $e^{3}=(1+{\omega})^{\frac{1}{2}}r sin{\theta}d{\phi}$.We shall addopt here a special set of tetrads where the connection 1-forms are given by ${\omega}^{a}_{b}=0$.This choice leads immediatly to the telaparallel condition
\begin{equation}
R^{a}_{bcd}({\Gamma})=0
\label{3}
\end{equation}
Since from the second Cartan equation for differential forms
\begin{equation}
R^{a}_{b}=R^{a}_{bcd}e^{c}{\wedge}e^{d}=d{\omega}^{a}_{b}+{\omega}^{a}_{e}{\wedge}{\omega}^{e}_{b}
\label{4}
\end{equation}
The torsion 2-form components $T^{a}=T^{a}_{bc}e^{b}{\wedge}e^{c}$, where $T^{a}_{bc}$ is the torsion tensor satisfy the first Cartan structure equation  
\begin{equation}
T^{a}=d{\omega}^{a}+{\omega}^{a}_{b}{\wedge}{\omega}^{b}
\label{5}
\end{equation}
Due to our choice of the connection 1-forms and (\ref{5}) makes the torsion 1-form to reduce to $T^{a}=d{\omega}^{a}$.Thus this expression yields some constraints on our metric which we shall pass to examine.From the above basis 1-forms we obtain the following torsion 2-forms 
\begin{equation}
T^{0}=d^{2}t=d^{2}r=0
\label{6}
\end{equation}
and the non-trivial components are
\begin{equation}
T^{2}=de^{2}=[(1+{\omega})^{\frac{1}{2}}+\frac{r{\omega}'}{2(1+{\omega})^{\frac{1}{2}}}]dr{\wedge}d{\theta}+[\frac{r\dot{\omega}}{{(1+{\omega})}^{\frac{1}{2}}}]dt{\wedge}d{\theta}
\label{7}
\end{equation}
and
\begin{equation}
T^{\phi}=[(1+{\omega})^{\frac{1}{2}}+\frac{r{\omega}'}{2(1+{\omega})^{\frac{1}{2}}}]sin{\theta}dr{\wedge}d{\phi}+[\frac{r\dot{\omega}}{(1+{\omega})^{\frac{1}{2}}}]dt{\wedge}d{\phi}+r(1+{\omega})^{\frac{1}{2}}cos{\theta}dr{\wedge}d{\phi}
\label{8}
\end{equation}
We solve the system for the following choice of torsion component
\begin{equation}
T^{\theta}=K dt{\wedge}d{\theta}
\label{9}
\end{equation}
where K is an arbitrary constant.This ansatz leads to the following system of partial diferential equations
\begin{equation}
(1+{\omega})+{\frac{r}{2}}{\omega}'=0
\label{10}
\end{equation}
and
\begin{equation}
\frac{r\dot{\omega}}{{(1+{\omega})}^{\frac{1}{2}}}=K
\label{11}
\end{equation}
Solution of this last equation yields
\begin{equation}
(1+{\omega})=\frac{K^{2}t^{2}}{{r}^{2}}
\label{12}
\end{equation}
This equation is compatible with the other differential equation.The remaining torsion component is simply
\begin{equation}
T^{\phi}=Ksin{\theta}dt{\wedge}d{\phi}
\label{13}
\end{equation}
Thus our solution ${\omega}=-(1-\frac{K^{2}t^{2}}{r^{2}})$ leads to the texture space-time
\begin{equation}
ds^{2}=dt^{2}-dr^{2}-\frac{K^{2}t^{2}}{r^{2}}r^{2}d{\Omega}^{2}
\label{14}
\end{equation}
One notes that in the light cone limit $t=r$ this metric reduces to
\begin{equation}
ds^{2}=dt^{2}-dr^{2}-K^{2}r^{2}d{\Omega}^{2}
\label{15}
\end{equation}
This metric is not globally Euclidean but in fact can be reduced to aflat metric if one considers the coordinate transformations ${\theta}'=\frac{\theta}{K}$ and ${\phi}'=\frac{\phi}{K}$ which shows that this geometry is not globally Euclidean,since the angles ${\theta}$ and ${\phi}$ varies in the range $0<{\theta}'<\frac{2{\pi}}{K}$ and the same for the angle ${\phi}$.Therefore the deficit angles depends both on torsion scalar $K$ as happens in the cosmic string case.Finally let us investigate the geodesics for the metric (\ref{1}) above.The geodesics equations are
\begin{equation}
\ddot{t}=-\frac{1}{2}r^{2}\dot{\theta}{\partial}_{t}{\omega}
\label{16}
\end{equation}
and the radial acceleration now is given by
\begin{equation}
\ddot{r}=\frac{1}{2}{\dot{\theta}^{2}}{\partial}_{r}[r^{2}(1+{\omega})]
\label{17}
\end{equation}
and finally 
\begin{equation}
\dot{[{\dot{\theta}^{2}}r^{2}(1+{\omega})]}=0
\label{18}
\end{equation}
At the light cone limit we have after some straightfoward algebra that the geodesics equations reduce to the equation $\ddot{r}=\frac{CK^{2}}{r^{3}}$ which shows that the acceleration of particles around the teleparallel texture suffer a torsion force which is inversely proportional to $r^{3}$.Just to compare our solution with the Notzold one we could say that at late times $t>>r$ his solution behaves as ${\omega}=\frac{t}{r}$ while our solution contains second order terms in the $\frac{t}{r}$ factor.This may be physically interpreted in terms of the weakness of the torsion force.Cosmological properties of telaparallel textures maybe studied elsewhere.
\section*{Acknowledgements}
I am very much indebt to Professors Christoffer Koehler and Patricio Letelier for helpful discussions
on the subject of this paper.Thanks are also due to Universidade do Estado do Rio de Janeiro
(UERJ) for financial Support.

\end{document}